\documentclass[twocolumn,notitlepage,prl,amsmath,amssymb,superscriptaddress,showpacs,floatfix]{revtex4-1}
\usepackage{hyperref}
\usepackage{mathtools}
\usepackage[usenames,dvipsnames]{color}
\usepackage[english]{babel}
\usepackage{graphicx}
\usepackage{epstopdf}
\usepackage{xcolor}
\usepackage{tabularx}

\begin{document}

\title{Entropy in the non-Fermi-liquid regime of the doped $2d$ Hubbard model}

\author{Connor Lenihan}
\affiliation{Department of Physics, King's College London, Strand, London WC2R 2LS, UK}

\author{Aaram J. Kim}
\email{aaram.kim@kcl.ac.uk}
\affiliation{Department of Physics, King's College London, Strand, London WC2R 2LS, UK}

\author{Fedor \v{S}imkovic IV.}
\affiliation{Department of Physics, King's College London, Strand, London WC2R 2LS, UK}
\affiliation{Centre de Physique Th\'eorique, \'Ecole Polytechnique,
CNRS, Universit\'e Paris-Saclay, 91128 Palaiseau, France} 
\affiliation{Coll\`{e}ge de France, 11 place Marcelin Berthelot, 75005 Paris, France}

\author{Evgeny Kozik}
\email{evgeny.kozik@kcl.ac.uk}
\affiliation{Department of Physics, King's College London, Strand, London WC2R 2LS, UK}
%

\begin{abstract}
We study thermodynamic properties of the doped Hubbard model on the square lattice in the regime of strong charge and spin fluctuations at low temperatures near the metal-to-insulator crossover and obtain results with controlled accuracy using the diagrammatic Monte Carlo method directly in the thermodynamic limit. The behavior of the entropy reveals a non-Fermi-liquid state at sufficiently high interactions near half-filling: a maximum in the entropy at non-zero doping develops as the coupling strength is increased, along with an inflection point, evidencing a metal to non-Fermi-liquid crossover. The specific heat exhibits additional distinctive features of a non-Fermi-liquid state. 
Measurements of the entropy can therefore be used as a probe of the state of the system in quantum simulation experiments with ultracold atoms in optical lattices.
\end{abstract}


\maketitle

The $2d$ Hubbard model~\cite{hubbard1963electron} is one of the foremost models of strongly interacting electrons, whose Hamiltonian is written as 
\begin{equation}
	\mathcal{H} = -  t\sum_{\left<ij\right>\sigma}\left(\hat{c}^{\dagger}_{i\sigma} \hat{c}^{}_{j\sigma} + \text{H.c.}\right)
+ U \sum_{i} \hat{n}_{i\uparrow} \hat{n}_{i \downarrow} - \mu \sum_{i\sigma} \hat{n}_{i\sigma}.
    \label{Hubbard}
\end{equation}
Here, $t$ is the nearest-neighbor hopping amplitude, $U$ is the on-site Coulomb interaction, $\mu$ is the chemical potential,
$\hat{c}^{\dagger}_{i\sigma}$ ($\hat{c}^{}_{i\sigma}$) creates (annihilates) a fermion with the spin $\sigma$ on the site $i$, and $\hat{n}_{i\sigma}=\hat{c}^{\dagger}_{i\sigma} \hat{c}^{}_{i\sigma}$. 
In spite of its seeming simplicity, it is widely believed that the Hubbard model captures the rich phenomenology of cuprate high-$T_c$ superconductors~\cite{anderson1997theory,Scalapino2012}. Nonetheless, after decades of extensive studies, much of the intriguing physics harbored by the Hubbard model remains to be uncovered and definitively described~\cite{LeBlanc2015}.

Among different properties of a thermodynamic system, the entropy has a special place, capturing the temperature dependence of several local thermodynamic observables via Maxwell relations. 
It is also where quantum many-body physics meets information theory~\cite{Islam2015,Cocchi2016,Cocchi2017,Walsh2019,Walsh2019a}. In previous studies, the entropy and some of the equations of state of the $2d$ Hubbard model have been computed by various state-of-the-art numerical algorithms, such as the determinant quantum Monte Carlo (DQMC)~\cite{Duffy1997,Paiva2001,Paiva2010, Gorelik2012}, dynamical cluster approximation (DCA)~\cite{LeBlanc2013,Galanakis2011}, numerical linked cluster expansion (NLCE)~\cite{Khatami2011,Khatami2012}, variational cluster approximation (VCA)~\cite{Seki2018}, and finite-temperature Lanczos method (FTLM)~\cite{Bonca2003}.
These results have substantially advanced our understanding of thermodynamics of the $2d$ Hubbard model, having led to a quantitatively reliable picture at high temperatures. 

A particular challenge has been the problem of how the metallic character in the $2d$ Hubbard model at low temperatures is destroyed by developing correlations~\cite{georges:1996, schafer:2015, rohringer2016, Fedor2020, kim2020, schaefer2020tracking}. 
It is central to understanding the role and nature of correlations that can potentially drive high-temperature superconductivity at appropriate doping~\cite{anderson1997theory,georges:1996,Scalapino2012}. 
It has been recently demonstrated~\cite{Fedor2020, kim2020} that the insulating behavior at half-filling (the average number of particles per lattice site $n=1$) emerges smoothly as $U$ is increased starting from a metallic state due to extending antiferromagnetic (AFM) correlations, while the charge correlation function reveals a correlation hole~\cite{kim2020}. Such a crossover is extended in parameter space, involving a transitional non-Fermi-liquid (NFL) state with a partially-gapped Fermi surface~\cite{Fedor2020}. 
This picture has proven difficult to capture even qualitatively with finite-size methods due to the long-range nature of correlations despite the absence of the fermionic sign problem at half-filling~\cite{schafer:2015, Fedor2020, kim2020, schaefer2020tracking}. 

On the other hand, experiments with ultracold atoms in optical lattices~\cite{Bloch2012,Lewenstein:2007hr,Greif:2015bg,Hart2015, Cheuk:2016kq,Mazurenko:2017ec,Brown:2017dy, Nichols:2019iq} accurately realize the model (\ref{Hubbard}) in a physical system and offer a route towards exploring its properties over regimes not accessible to theoretical methods. 
Recent experiments have approached the range of temperatures $T \lesssim 0.25t$ of the metal-to-insulator crossover: they demonstrate development of long-range AFM correlations at $T \sim 0.2t$~\cite{Mazurenko:2017ec} and their disappearance with doping, while {\it in situ} measurements of the equation of state $n(\mu,T)$~\cite{Cocchi2016,Cocchi2017} enable accurate determination of the entropy. However, reliable theoretical predictions for thermodynamics of the doped $2d$ Hubbard model in this regime are missing. Not only are controlled predictions crucial for validation of upcoming experiments, but also experimental determination of equations of state and, ultimately the phase diagram of the model (\ref{Hubbard}), requires reliable thermometry, which in these isolated systems is only enabled by theoretical input. 

\begin{figure*}[htbp]
\centering

\includegraphics[width=1.0\textwidth]{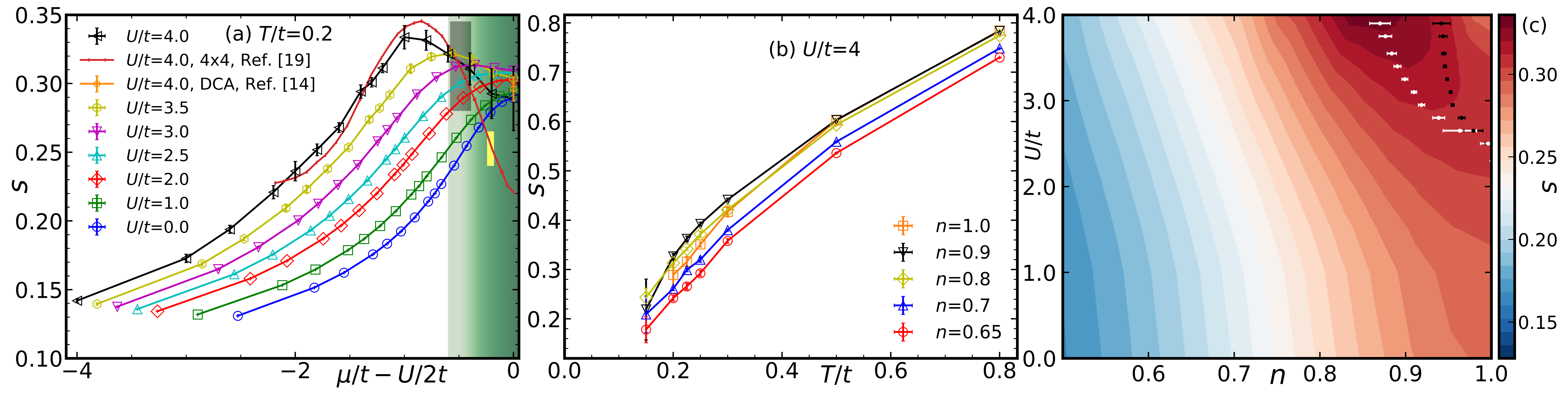}
\caption{ 
	(a) Entropy as a function of the chemical potential for different $U$ at $T=0.2t$ [For $U=4t$ the gray shading marks the regime where $\partial D /\partial T>0$, which contains the region marked by the green shading where the state of the system is of a NFL character, $\partial \kappa / \partial T >0$; the dark gray bar represents the location of the crossover $\mu_{\mathrm{ch}}$ with its error bar, the yellow bar being its counterpart for the FTLM data on a $4 \times 4$ lattice of Ref.~\cite{Bonca2003}]; extrapolated to the TDL DCA result~\cite{LeBlanc2013} at $U=4t$ verifies our calculation at half-filling. (b) Entropy as a function of the temperature for various densities at the fixed interaction strength $U=4t$.
	(c) Color map of the entropy in the $n$-$U$ plane for $T=0.2t$, where the black points pinpoint the location of the crossover from the metallic to NFL regime, $n(\mu_{\mathrm{ch}},U)$, and the white points mark the entropy maximum.}
\label{fig:entropy}
\end{figure*}

Here we study the entropy and thermodynamics of the doped Hubbard model on the square lattice in the highly non-trivial correlated regime where the system experiences a crossover from metallic Fermi-liquid (FL) to NFL behavior in the presence of competing energy scales and strong correlations. We obtain results with controlled accuracy directly in the thermodynamic limit (TDL) by the diagrammatic Monte Carlo approach, in which the coefficients of the perturbative expansion for a particular observable in powers of the coupling $U$ are computed by a numerically exact stochastic sampling of all contributing Feynman diagrams~\cite{VanHoucke:2010ky,Kozik:2010fla}. More specifically, we use the connected determinant (CDet) diagrammatic Monte Carlo algorithm~\cite{Rossi2016det} and the approach of Ref.~\cite{Fedor2019} for controlled evaluation of observables from their diagrammatic series in the strongly correlated regime, previously employed in Refs.~\cite{Fedor2020, kim2020}. The behavior of entropy as a function of doping and interaction strength contains ample information about the state of the system, allowing, in particular, to pinpoint the FL-NFL crossover in the charge sector by its relation to double occupancy and compressibility. At the same time, the specific heat shows signatures of the crossover in the AFM channel and suppression of the density of states in the partially gapped NFL at half-filling. 
Our results suggest that the most basic techniques of cold atom experiments, such as adiabatic loading of a Fermi gas with known entropy in the optical lattice~\cite{PhysRevLett.115.260401} and determination of entropy from the equation of state~\cite{Cocchi2016,Cocchi2017}, can be used for detecting the state of the system in this regime, while our data for the temperature dependence of entropy and $n(\mu,T)$ provide a basis for reliable thermometry. The maximum of entropy at a particular doping in the correlated regime can indicate the vicinity of phase separation~\cite{Galanakis2011,macridin2006phase, aichhorn2007phase, chang2008spatially, sorella2015finite, zheng2016ground} or favor high-temperature superconductivity, as observed in cuprates~\cite{Tallon2004, Storey2007, Michon2019}.

We obtain the entropy of the system using the formula:
\begin{equation}
s = \frac{1}{T}(\mathcal{E}_K +UD-\mu n + P)~, \label{entropy}
\end{equation}
where $\mathcal{E}_{K}$ is the kinetic energy, $D$ is the double occupancy, $P$ is the negative grand potential density, equal to the pressure for a homogeneous system, and all extensive quantities are defined per lattice site. Having computed the series coefficients for $\mathcal{E}_K$, $D$, $n$, and $P$, we evaluate these observables using the procedure developed in Ref.~\cite{Fedor2019} (for more details of the method and its limitations, see~\cite{SM}) and compute the entropy with Eq.~(\ref{entropy}) directly in the TDL without numerical integration or a fitting procedure.

Figure~\ref{fig:entropy}(a) shows the entropy as a function of chemical potential (shifted by the half-filling value $U/2$) for a fixed temperature $T=0.2t$ (results at $T=0.3t$ are also presented in~\cite{SM}), which is within the range $T \lesssim 0.25t$, where quasiparticle properties can be defined near half-filling~\cite{Fedor2020} (and thus one can speak of FL and NFL states). This temperature has also become accessible in state-of-the-art experiments. At half-filling, $s(\mu=U/2)$ for $U=4t$ is in perfect agreement with the extrapolated to the TDL DCA result~\cite{LeBlanc2013}. The key observation is that $s(\mu)$ changes its shape qualitatively as $U$ is increased~\cite{Bonca2003,Khatami2011}, which can be understood in the following way. 
Due to the particle-hole symmetry of the Hamiltonian, i.e. the symmetry of Fig.~\ref{fig:entropy}(a) w.r.t. reflection about the $\mu=U/2$ axis, and the requirement that $s(\mu)$ is smooth, $\partial s /\partial \mu =0$ at $\mu=U/2$. [By the Maxwell relation $\partial s /\partial \mu = \partial n / \partial T$ it translates to the well-known fact that $n=1$ at $\mu=U/2$ at any temperature.] 
Thus, the drift of the maximum of $s(\mu)$ away from half-filling at $U$ values above a certain threshold $U_{\text{ch}}$ must be marked by the appearance of an inflection point at $\mu_{\text{ch}}(U)$ close to half-filling, defined by $\partial^2 s(\mu_{\text{ch}}) / \partial \mu^2 =0$. 
By the Maxwell relation $\partial^2 s/ \partial \mu^2 = \partial \kappa / \partial T$, where $\kappa = \partial n/\partial \mu$ is the compressibility, this inflection point first appears when $\partial \kappa / \partial T$ at half-filling changes sign, signaling the crossover from metallic to insulating behavior: a half-filled metal is characterized by $\partial \kappa / \partial T < 0$, while an insulator features $\partial \kappa / \partial T > 0$~\cite{kim2020}. 
At $T=0.2t$, this crossover was found in Ref.~\cite{kim2020} to happen at $U_{\text{ch}} \approx 2.5t$, which shows in Fig.~\ref{fig:entropy}(a) by the maximum moving away from $\mu=U/2$ between the curves for $U=2t$ and $U=3t$. 
As $U$ is increased further above $U_{\text{ch}}$, the inflection point at $\mu_{\text{ch}}(U)$ moves away from half-filling, implying that the state of the system is of a NFL character in the growing range of dopings such that $\mu > \mu_{\text{ch}}(U)$. The NFL regime for $U=4t$ is marked in Fig.~\ref{fig:entropy}(a) by the green shading and the location of the crossover by the black points in Fig.~\ref{fig:entropy}(c).

\begin{figure}[htbp]
\begin{center}
\includegraphics[width=1.0\columnwidth]{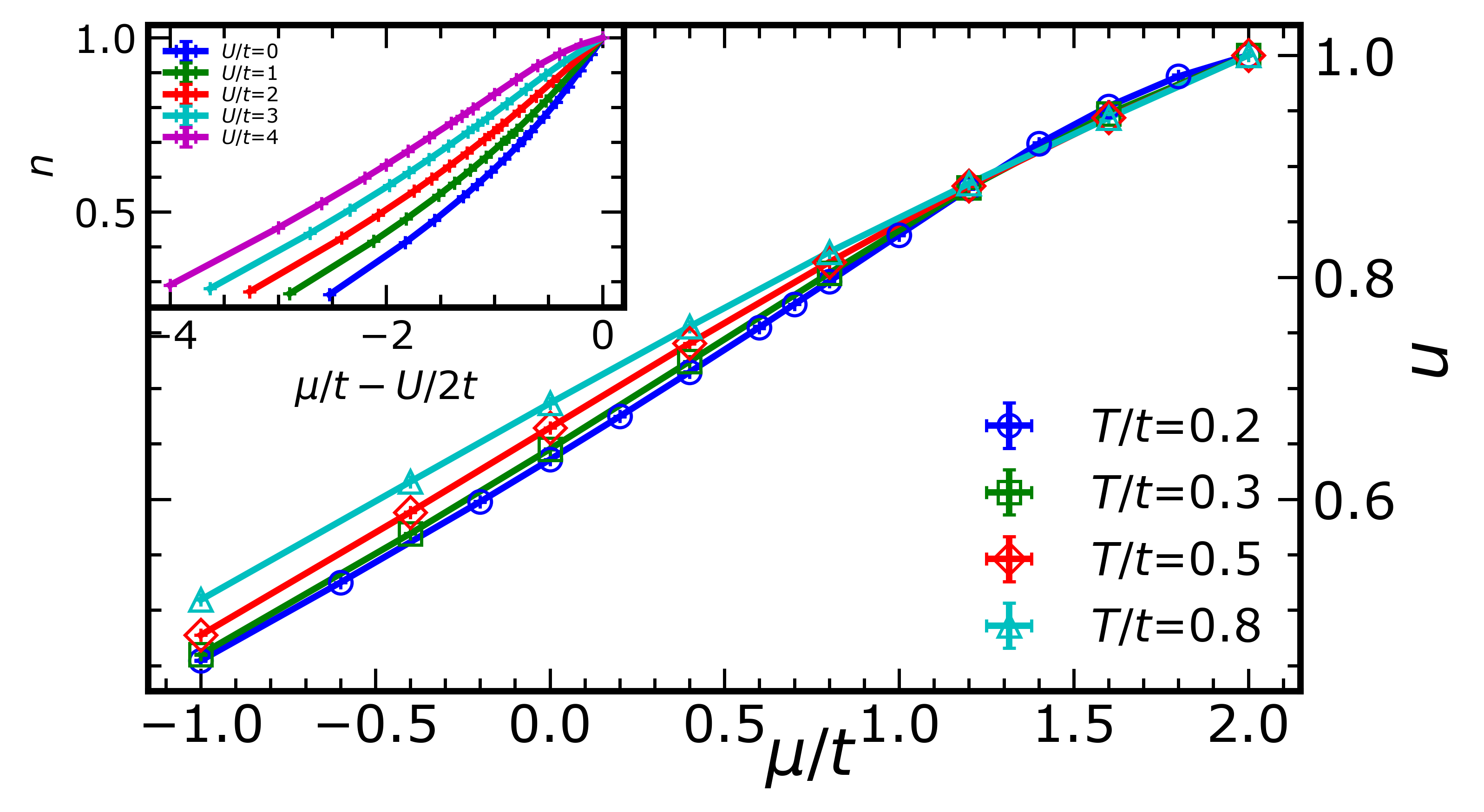}

\end{center}
\caption{ 
	Density as a function of the chemical potential at different temperatures for $U=4t$.
	Inset shows the density as a function of $\mu-U/2$ for various $U$ at $T=0.2t$~.
}
\label{fig:density_mu}
\end{figure}

The presence of the entropy maximum at $\mu_{\text{max}}$ away from half-filling can be detected via $\partial s /\partial \mu = \partial n / \partial T=0$ in the equation of state $n(\mu,T)$ for a given interaction strength.
The $n(\mu)$ curves for close enough temperatures must cross at the location of the maximum of $s(\mu)$, as seen for $U=4t$ in Fig.~\ref{fig:density_mu}. 
Figure~\ref{fig:entropy}(c) combines the raw data points of $s(\mu)$ in Fig.~\ref{fig:entropy}(a) with the equation of state (Fig.~\ref{fig:density_mu}), and shows the entropy map in the $n$-$U$ plane for a fixed temperature $T=0.2t$~. 
Along the trajectory $\mu_{\text{max}}(U)$, the entropy monotonically increases as a function of $U$. 

Figure~\ref{fig:entropy}(a) compares the TDL $s(\mu)$ with the numerically exact FTLM data for a $4 \times 4$ lattice~\cite{Bonca2003} at $U=4t$~\footnote{The original FTLM data are available for $s(n)$, the $s(\mu)$ curve being obtained using the equation of state in Fig.~\ref{fig:density_mu}.}. There is good agreement within error bars at $\mu \lesssim \mu_{\text{max}}$, but, inside the NFL regime, the finite-size data drop significantly below the TDL curve with the discrepancy reaching $\sim 25\%$. The underestimation of entropy could be attributed to fluctuations beyond a few lattice sites, likely due to paramagnons, as found at half-filling~\cite{Fedor2020, kim2020}. Thus, spatial correlations remain important throughout the NFL regime, which appears in a significantly reduced range of doping on the $4 \times 4$ lattice. Interestingly, the density $n^* \approx 0.88$ of the entropy maximum at $U=4t$ shows no significant change from the finite size system at $U=4t$ \cite{Bonca2003}, or $U=6t$ \cite{Galanakis2011} and is already close to that reported for the $t$-$J$ model ($\approx 0.85$)~\cite{Jaklic1996}. Finally, Fig.~\ref{fig:entropy}(b), where $s(T)$ is plotted at several densities, can be used for thermometry in experiments where a weakly-interacting gas of known entropy is loaded in the optical lattice adiabatically~\cite{PhysRevLett.115.260401}.

At a certain $\mu_{\text{pot}}(U)$, the $s(\mu)$ curves corresponding to different $U$ values cross. By the Maxwell relation $\partial s /\partial U = - \partial D / \partial T$, at this point the temperature dependence of the potential energy $\mathcal{E}_P = U D$ changes sign:  $\partial D / \partial T > 0$ in the vicinity of half-filling for $\mu > \mu_{\text{pot}}(U)$, meaning that in this regime, double occupancy drops with cooling. At the same time, since $\partial n/\partial T$ is negative for $\mu > \mu_{\text{max}}$, this ensures the development of the local magnetic moment $\langle s_z^2 \rangle = n/4 - D/2$ upon cooling towards the quasi-antiferromagnet. 
Curiously, at half-filling and fixed $T$, development of the local moment~\cite{Seki2018} was found in Ref.~\citep{kim2020} to start at a larger value of $U$ than the $U_{\text{ch}}$ needed for the compressibility to acquire its insulating character. This was attributed to the substantial enhancement of $\partial \kappa / \partial T$ due to the temperature dependence of non-local charge fluctuations. The behavior of entropy in Fig.~\ref{fig:entropy}(a) shows that, with growing interactions, $\mu_{\text{pot}}(U)$ decreases faster than $\mu_{\text{ch}}(U)$ from $U/2$, eventually passing it (marginally, given the error bars), so that at $U=4t$ the range of dopings for which  $\partial D / \partial T > 0$ is somewhat wider than the NFL region defined by $\mu > \mu_{\text{ch}}$.
In Fig.~\ref{fig:entropy}(a) at $U=4t$, the region where the double occupancy decreases with cooling, $\partial D / \partial T > 0$, is denoted by the gray shading, which also contains the NFL regime.

\begin{figure}[htbp]
\includegraphics[width=1.\columnwidth]{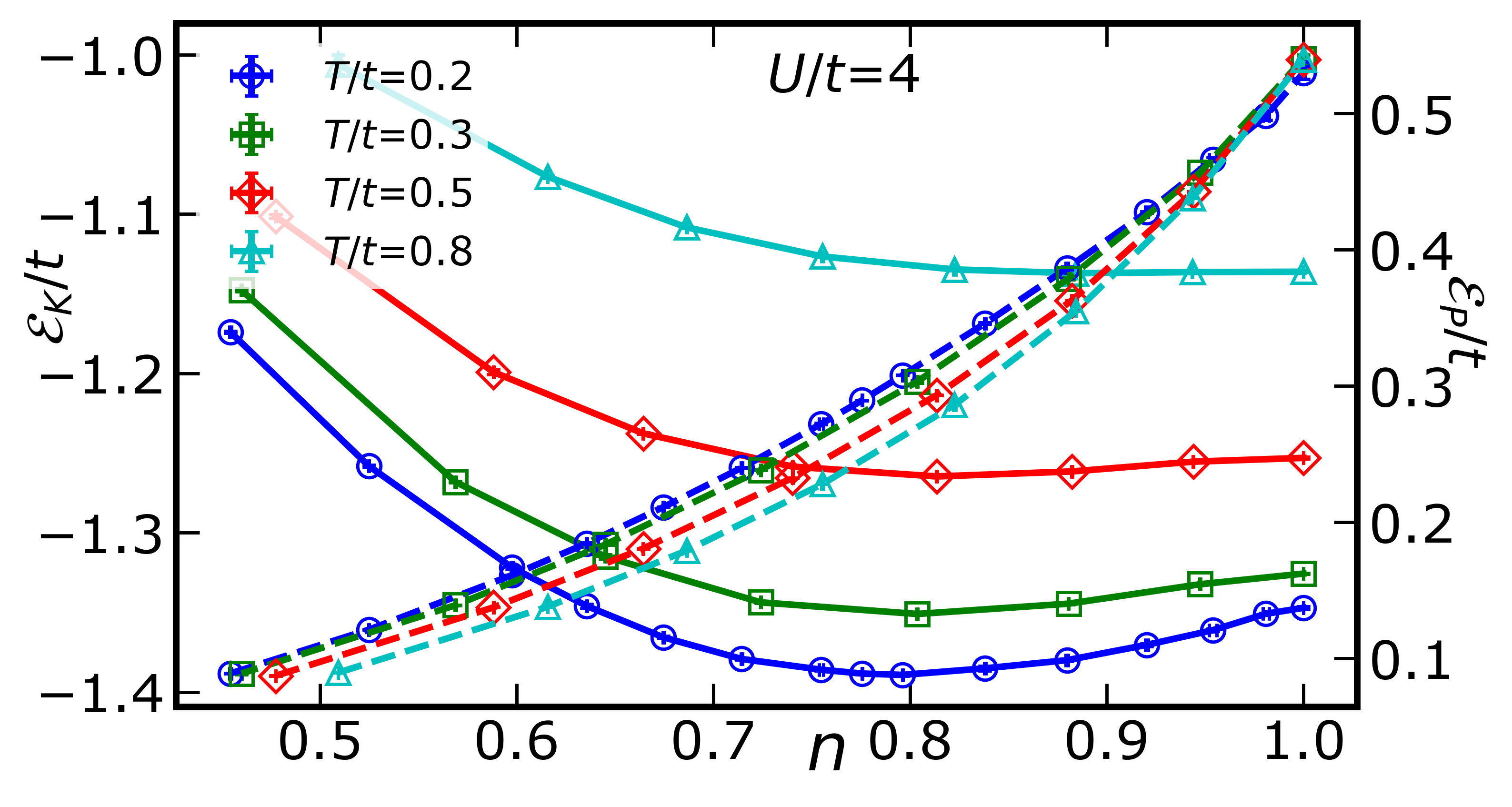}
\caption{
	The kinetic (solid lines) and potential energy densities (dashed lines) plotted against density for various temperature at fixed $U=4t$. 
}
\label{fig:ke_pe}
\end{figure}
In Ref. \cite{Arzhnikov2012} magnetic phase separation - with a critical point at $T\approx0.23t$, $n\approx0.88$ - was predicted using a mean-field theory with transverse magnetic fluctuations. Figure \ref{fig:density_mu} shows no sign of the divergent derivative  of $n(\mu)$ characteristic of the transition~\cite{Galanakis2011}, ruling out phase separation for $T\geq0.2t$. However, the entropy maximum and the rising derivative in $n(\mu)$ at $\mu_{\text{max}}$ are consistent with phase separation near $\mu_{\text{max}}$ at considerably lower temperatures.

Figure~\ref{fig:ke_pe} shows the behavior of the kinetic $\mathcal{E}_K$ and potential $\mathcal{E}_P$ energies as a function of density for various temperatures at $U=4t$. 
While at high temperatures $\mathcal{E}_K(n)$ monotonically decreases on approach to half-filling, a minimum appears at $T \lesssim 0.5t$, which propagates from half-filling to lower densities upon cooling. 
The rise of $\mathcal{E}_K(n)$ on approach to $n=1$ is a signature of increasing localization in the NFL regime. 
In contrast, $\mathcal{E}_P(n)$ is a monotonically increasing function for all $T$. 
However, the curves for neighboring temperatures cross at the point where $\partial D/\partial T =0$ ($= - \partial s/\partial U$), corresponding to the location of the boundary of the gray region in Fig.~\ref{fig:entropy}(a).

\begin{figure}[htbp]
\centering
\includegraphics[width=1.0\columnwidth]{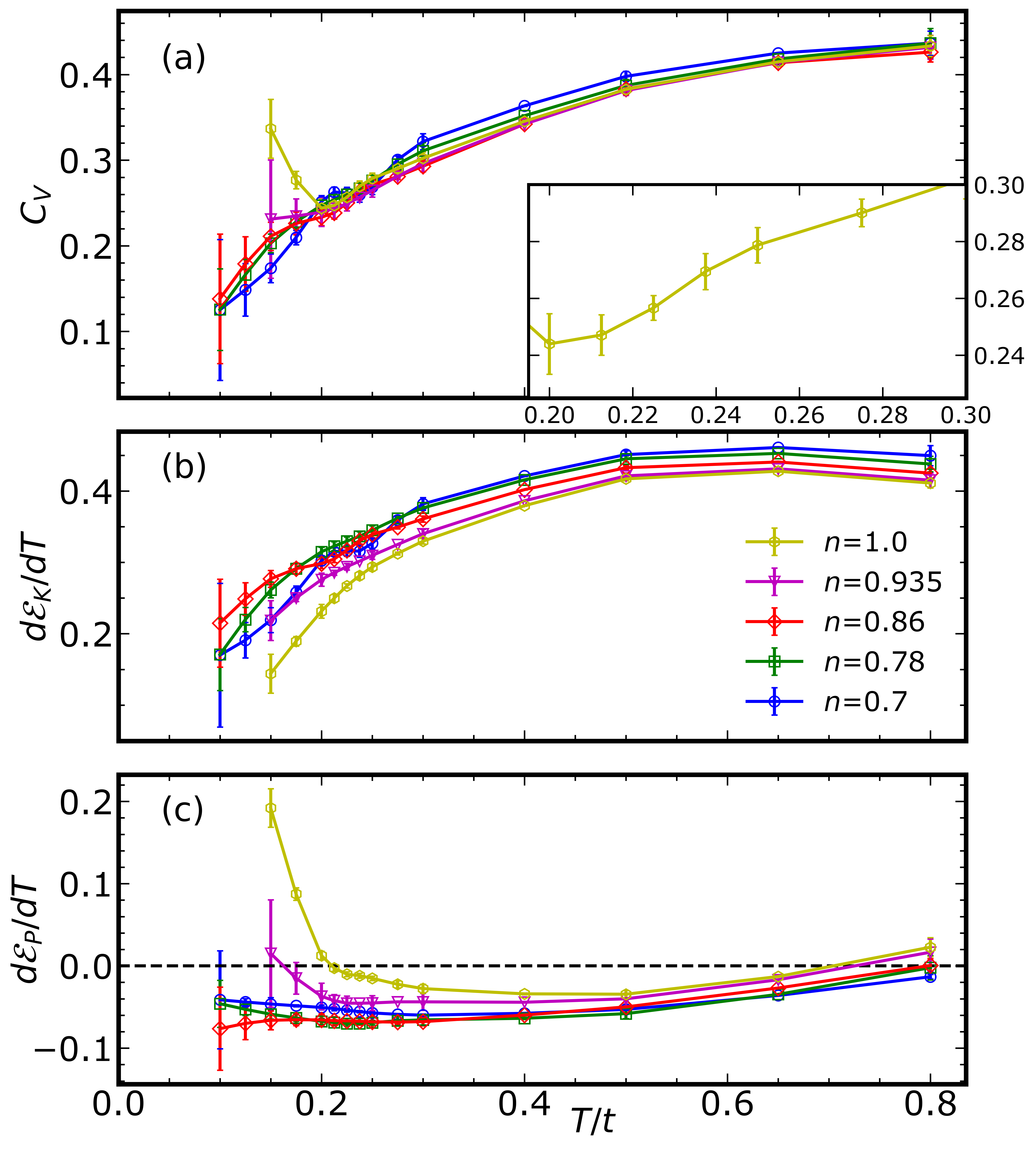}
\caption{
	(a) The specific heat as a function of temperature for different densities at $U=3t$, the inset zooming on a small shoulder at $n=1$. 
	(b) The kinetic and (c) the potential energy contribution to the specific heat as functions of temperature.
}
\label{fig:Cv} 
\end{figure}

Numerical differentiation of the total energy $\mathcal{E}$ allows us to obtain the specific heat $C_V$, which exhibits several signatures of the rich metal-to-NFL crossover physics. 
Figure~\ref{fig:Cv} shows $C_V(T)$ for different densities at $U=3t$. 
At half-filling, $C_V(T)$ develops a sharp upturn, which we anticipate based on the data at other parameters ~\cite{Duffy1997,Paiva2001,Seki2018} to have a maximum at $T \lesssim 0.15t$ (we are not able to reliably evaluate $C_V$ at temperatures lower than $T=0.15t$ at $n=1$). 
Such a maximum is generically due to extending AFM correlations and is also seen, e.g., at the N\'eel transition in $3d$~\cite{PhysRevB.87.205102}. 
In $2d$ (where a phase transition at $n=1$, $T>0$ is forbidden), it can be viewed as a marker for the crossover to the quasi-AFM state with a large correlation length of more that $10$ lattice sites, which is predicted in Ref.~\citep{kim2020} to happen at $T \sim 0.15t$ for $U=3t$, in consistency with our results. 
Earlier finite-size DQMC and VCA results at half-filling appear to capture the location of the peak and a broad maximum at high temperatures~\cite{Duffy1997,Paiva2001,Seki2018}. It is interesting, however, that $C_V(T)$ at half-filling also exhibits a small shoulder at $T \approx 0.25t$, where according to Ref.~\cite{Fedor2020} the single-particle gap first opens at the antinodal point of the Brillouin zone. The accelerated decrease of $C_V(T)$ at temperatures immediately below $T = 0.25t$ is consistent with the picture of the gap proliferating along the Fermi surface before the AFM correlations become appreciable at $T \approx 0.2t$ and $C_V(T)$ starts rising sharply. 

As the system is doped, the AFM peak is dramatically suppressed already at $n \sim 0.935$ and completely disappears at $n \lesssim 0.86$ down to the lowest accessible temperatures. At large dopings, the $C_V(T)$ curve becomes essentially featureless, suggesting that it approaches its low-temperature $C_V(T) \propto T$ FL asymptote. However, at the specific density $n=0.7$, which is well below the location of the entropy peak, and at which the system is expected to be a good FL at low enough $T$, we detect a sizeable bump in $C_V(T)$ at $T \approx 0.2t$, which appears to be robust w.r.t. various  numerical differentiation schemes. 

Splitting the specific heat into the kinetic ($\partial\mathcal{E}_{K}/\partial T$, Fig.~\ref{fig:Cv}(b)) and potential ($\partial \mathcal{E}_{P}/\partial T$, Fig.~\ref{fig:Cv}(c)) energy contributions~\cite{Paiva2001}, sheds light on the origin of the features of $C_V(T)$: The low temperature upturn close to half-filling is entirely due to the potential energy, which is consistent with the metal-to-insulator crossover being driven by Slater physics~\cite{Rev:1951ib}, whereas the increase of $C_V$ at higher temperatures is due to $\mathcal{E}_{K}$. 
As the system is doped, the AFM peak in $\mathcal{E}_{P}$ gradually disappears around densities where the entropy is maximal. 
Further doping extends the range of temperatures where $\partial\mathcal{E}_P/\partial T<0$, in consistency with the increasing entropy with $U$, seen in the large doping regime in Fig.~\ref{fig:entropy}(a).

In conclusion, we have demonstrated by controlled calculations in the TDL that entropy and specific heat contain ample information to characterize the state of the doped $2d$ Hubbard model, with respect to both its charge and spin properties, and observe the crossover from metallic to NFL behavior at low enough temperatures where quasiparticles become meaningful. This regime is already within reach of current experiments with ultracold atoms in optical lattices. 

The maximum in $s(n)$ that appears at a non-zero doping is entirely due to correlations, and its emergence necessarily requires a NFL regime developing in the vicinity of half-filling. A sharp maximum of entropy is expected, e.g., above the critical end-point of the phase separation line, which was suggested to emerge at a lower $T$ in this regime of parameters~\cite{Galanakis2011,macridin2006phase, aichhorn2007phase, chang2008spatially, sorella2015finite, zheng2016ground}. By the condition $\partial^2 s/ \partial \mu^2 = \partial \kappa /\partial T$, it could stem from the divergence of $\kappa$ on approach to the critical point. On the other hand, an entropy maximum corresponds to a high density of many-body states available for scattering, while thermodynamic observables become NFL-like only closer to half-filling than the location of the maximum. Thus, the entropy maximum could favor a Fermi surface instability, such as Cooper pairing, which could occur at a relatively high temperature in its vicinity. In fact, the maximum of the electronic entropy has been observed to correlate with the highest superconducting $T_c$ in cuprates~\cite{Tallon2004, Storey2007, Michon2019}.

Here we focused on the interaction range $U \lesssim 4t$, to which the metal-to-insulator crossover is confined at half-filling: at larger couplings cooling the system brings it directly in the insulating regime~\cite{Fedor2020}. 
In this range of $U$, correlations are also strongly non-local, making the formulation in the TDL a major advantage of our method. Although our technique is unable to produce reliable results at larger $U$, it is seen from Fig.~\ref{fig:entropy}(a) that the entropy maximum drifts toward larger dopings and keeps rising with increasing $U$ at these temperatures. 
From the behavior of magnetic correlations with doping~\cite{vsimkovic2017magnetic}, we expect that this trend continues until about $U \sim 6t$, where the entropy likely reaches its absolute maximum. At $U \gtrsim 6t$, the competition of superconducting and magnetic (stripe) phases in the ground state for $n \gtrsim 0.8$ has been recently studied using advanced tensor network and quantum Monte Carlo methods~\cite{Zheng2017stripe, Ido2018no_superconductivity, Qin2019absence_of_superconductivity}, where it was found that the stripes wipe out superconductivity. Our data rule out magnetic phase separation \cite{Arzhnikov2012} at $T\geq0.2t$ but are consistent with its existence at lower or even zero temperature \cite{Galanakis2011}, where magnetic fluctuations in the vicinity of the critical point could drive high-$T_c$ superconductivity, the regime of $4 \lesssim U/t \lesssim 6$, $n \sim 0.88$ being the best candidate for this scenario. 

\begin{acknowledgments}
This work was supported by EPSRC through the grant EP/P003052/1 and by the Simons Foundation as a part of the Simons Collaboration on the Many Electron Problem. We are grateful to the UK Materials and Molecular Modelling Hub for computational resources, which is partially funded by EPSRC (EP/P020194/1).
\end{acknowledgments}

\bibliography{refs}

\end{document}